\documentclass[conference]{IEEEtran}
\IEEEoverridecommandlockouts
\usepackage{cite}
\usepackage{amsmath,amssymb,amsfonts}
\usepackage{algorithmic}
\usepackage{graphicx}
\usepackage{textcomp}
\usepackage{xcolor}
\usepackage{multirow}
\usepackage{multicol}
\usepackage{caption} 
\usepackage{subcaption}
\usepackage{booktabs}
\usepackage{newfloat}
\usepackage{listings}

\def\BibTeX{{\rm B\kern-.05em{\sc i\kern-.025em b}\kern-.08em
    T\kern-.1667em\lower.7ex\hbox{E}\kern-.125emX}}
\begin{document}

\title{Tree-structured Auxiliary Online Knowledge Distillation\\
\thanks{This research is supported by  National Natural Science Foundation of China (Grant No. 6201101015) , Beijing Academy of Artificial Intelligence(BAAI), Natural Science Foundation of Guangdong Province (Grant No. 2021A1515012640), the Basic Research Fund of Shenzhen City (Grant No. JCYJ20210324120012033 and JCYJ20190813165003837), National Key R\&D Program of China (No. 2021ZD0112905)and Overseas Cooperation Research Fund of Tsinghua Shenzhen International Graduate School  (Grant No. HW2021008).}
}

\author{
\IEEEauthorblockN{1\textsuperscript{st} Wenye Lin}
\IEEEauthorblockA{\textit{Tsinghua Shenzhen International Graduate School} \\
\textit{Tsinghua University}\\
Shenzhen, China \\
lwy20@mails.tsinghua.edu.cn}
\and
\IEEEauthorblockN{2\textsuperscript{nd} Yangning Li}
\IEEEauthorblockA{\textit{Tsinghua Shenzhen International Graduate School} \\
\textit{Tsinghua University}\\
Shenzhen, China \\
liyn20@mails.tsinghua.edu.cn}
\and
\IEEEauthorblockN{3\textsuperscript{rd} Yifeng Ding}
\IEEEauthorblockA{\textit{Tsinghua Shenzhen International Graduate School} \\
\textit{Tsinghua University}\\
Shenzhen, China \\
dingyf20@mails.tsinghua.edu.cn}
\and
\IEEEauthorblockN{4\textsuperscript{th} Hai-Tao Zheng*\thanks{*Corresponding author.}}
\IEEEauthorblockA{\textit{Tsinghua Shenzhen International Graduate School} \\
\textit{Tsinghua University}\\
Shenzhen, China \\
zheng.haitao@sz.tsinghua.edu.cn}
}

\maketitle

\begin{abstract}
Traditional knowledge distillation adopts a two-stage training process in which a teacher model is pre-trained and then transfers the knowledge to a compact student model. To overcome the limitation, online knowledge distillation is proposed to perform one-stage distillation when the teacher is unavailable. Recent researches on online knowledge distillation mainly focus on the design of the distillation objective, including attention or gate mechanism. Instead, in this work, we focus on the design of the global architecture and propose Tree-Structured Auxiliary online knowledge distillation (TSA), which adds more parallel peers for layers close to the output hierarchically to strengthen the effect of knowledge distillation. Different branches construct different views of the inputs, which can be the source of the knowledge. The hierarchical structure implies that the knowledge transfers from general to task-specific with the growth of the layers. Extensive experiments on 3 computer vision and 4 natural language processing datasets show that our method achieves state-of-the-art performance without bells and whistles. To the best of our knowledge, we are the first to demonstrate the effectiveness of online knowledge distillation for machine translation tasks. 
\footnote{Code is available at https://github.com/Linwenye/Tree-Supervised.}
\end{abstract}


\section{Introduction}
Deep neural networks have led to a series of successes in computer vision and natural language processing \cite{lecun2015deep}. They show superiority in representing complex concepts due to the large size of parameters \cite{krizhevsky2012imagenet}. However, the cumbersome neural networks are computationally expensive, hindering their applications in real-world problems with limited resources. To this end, knowledge distillation \cite{hinton2015distilling} is proposed to transfer the knowledge from a large teacher model to a compact student model.

Traditional knowledge distillation requires a two-stage training process, in which a high-capacity teacher model is pre-trained in the first stage, and then transfers the knowledge of the teacher to a compact student model in the second stage \cite{hinton2015distilling,ba2014deep}. This two-stage process increases the pipeline complexity and training cost. To simplify the distillation procedure, online knowledge distillation \cite{zhang_deep_2018,lan2018knowledge} is proposed, which simultaneously trains a set of student models and distills their knowledge from each other in a peer-teaching manner. This approach requires a one-stage learning procedure and leverages peer network to provide the teacher knowledge. Recent researches on online knowledge distillation aim at improving the quality of the knowledge learned from peers. ONE \cite{lan2018knowledge} and CL-ILR \cite{song2018collaborative} introduces an ensemble teacher which gathers the students' knowledge with a gate mechanism. OKDDip \cite{chen2020online} further boosts the performance by maintaining the peer variety through an attention mechanism. These approaches mainly focus on the design of the distillation objective. 

\begin{figure*}
  \centering
  \includegraphics[width=13.5cm]{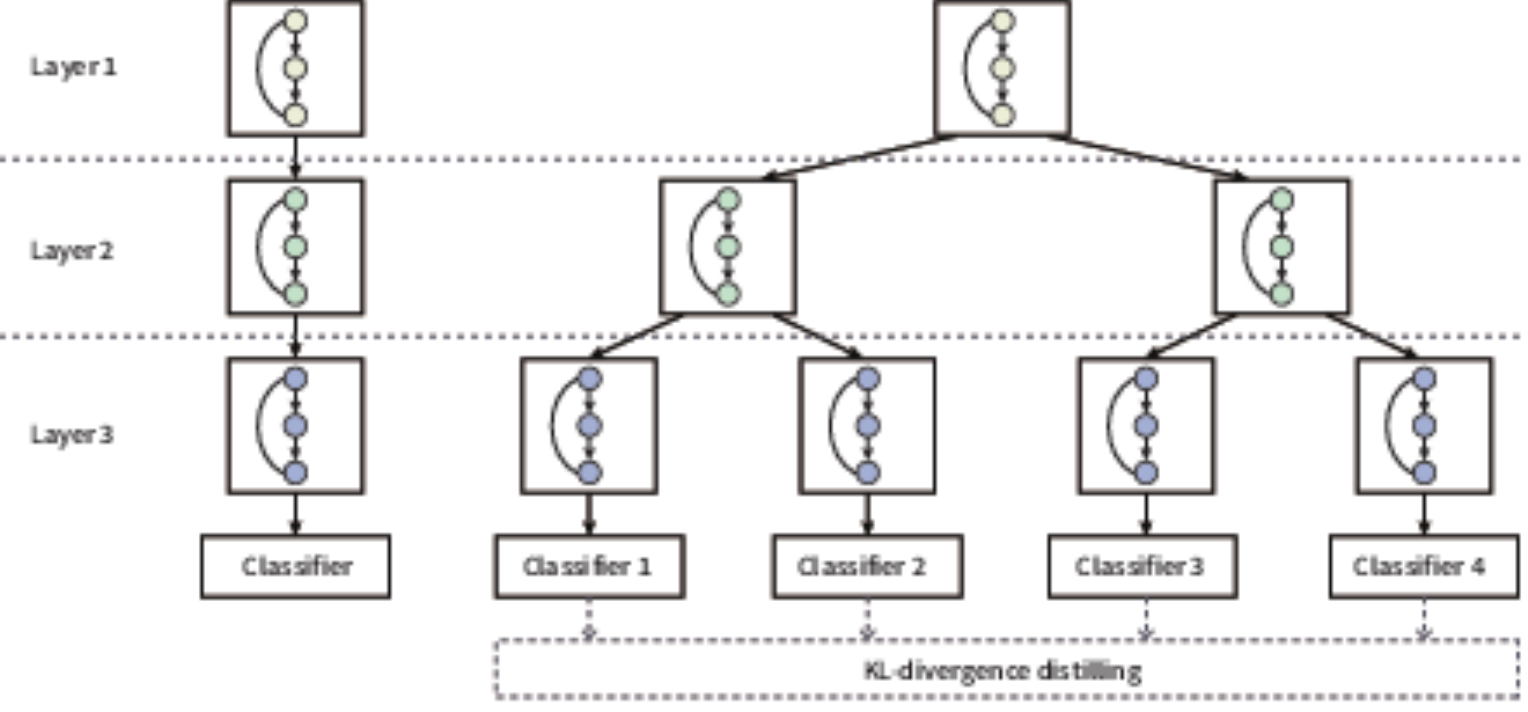}
  \caption{Deep tree-structured auxiliary online knowledge distillation example. On the left is an original network. On the right is training the network with TSA. In this example, we split the original network into three layers in a tree, and add replicas with different initialization to form a binary tree during training. Every classifier is trained to learn from both hard labels and soft targets from peers. At inference time, replicas are removed and the whole network recovers to the original architecture. With this tree manner, block on layer 1 is supervised by four classifiers, while block on layer 2 is supervised by two. Earlier layers are given more powerful ``instructions" from more classifiers. Because of different initialization, multiple branches provide various and abundant information to extract robust features for down-stream tasks.}
  \label{TSA example}

\end{figure*}

By contrast, in this work, we demonstrate that the design of the overall architecture for online knowledge distillation is another key factor for the performance of the student model. We adopt the original distillation objective as that in DML \cite{zhang_deep_2018} without bells and whistles, and show that the tree structure for online knowledge distillation is the key to our state-of-the-art performance. We propose a unified framework, the tree-structured auxiliary online knowledge distillation (TSA). During training, TSA hierarchically adds more auxiliary peers in later layers, which naturally forms a tree structure (See Fig. \ref{TSA example}). Different branches construct different views of the inputs, which can be the source of the knowledge. The hierarchical structure implies that the knowledge transfers from general to task-specific with the growth of the layers. Each branch of the tree is trained with joint objective loss terms: a conventional softmax cross-entropy loss to match the ground-truth hard labels, and a distillation loss that aligns each classifier's class posterior with the class probabilities of peers. The whole tree is trained together, with the sum of loss from every classifier as the overall loss. At the inference stage, all auxiliary modules are removed to keep the original architecture unchanged, thereby making the framework feasible to be put into deployment. Extensive experiments on both computer vision and natural language processing tasks show that our method is simple but effective and generalizes well to different domains.

Our main contributions are summarized as below:
\begin{enumerate}
\item We introduce tree-structured auxiliary online knowledge distillation, a unified framework for online knowledge distillation. The framework is simple but effective. We show that the tree structure is another key factor for the performance and provide a general perspective to get better understanding of online knowledge distillation. This could inspire new effective methods.
\item We implement two variants of TSA: a balanced tree and an unbalanced tree. Experiments show significant improvements in our method over vanilla training and existing online knowledge distillation methods. Typically, with online knowledge distillation of the balanced variant, all the networks gain an average of 3\% to 4\% improvement in accuracy on CIFAR-100, where MobileNetV3 and ResNet-18 reach the best accuracies ever reported for these architectures. On ImageNet, ResNet-34 obtains 74.97\% accuracy, which is 1.8\% above the vanilla one.
\item To the best of our knowledge, we are the first to demonstrate the effectiveness of online knowledge distillation for machine translation. On IWSLT translation tasks, we gain an average of 0.9 BLEU improvement over vanilla Transformer for three datasets. Experiment on large-scale translation task WMT'17  reveals that our method generalizes well to large datasets.
\end{enumerate}

\section{Related work}

\subsection{Knowledge Distillation}
Knowledge distillation is a ubiquitous technique for model compression and acceleration. It trains a smaller student model to imitate the output distribution of a larger teacher model via a $L_2$ loss on logits (i.e. before Softmax) \cite{ba2014deep} or cross-entropy \cite{hinton2015distilling}. Recently, many researches attempt to further improve the performance by discovering different knowledge from the teacher, such as intermediate representations \cite{romero2014fitnets}, attention maps \cite{zagoruyko2016paying}, and flow of solution process (FSP) \cite{yim2017gift}. However, the limitation to these methods is that they need a pre-trained teacher and conduct a two-stage distillation process.

\subsection{Deeply Supervised Nets} Deeply supervised nets (DSN) \cite{lee2015deeply} emphasizes the direct supervision on intermediate layer by introducing classifier to each layer, aiming at extracting more discriminative features. Self-distill \cite{zhang2019your} further facilitates this kind of supervision on hidden layers through distilling knowledge from the later classifier to earlier ones. However, a deeply supervised method may downgrade the performance of deep nets. Features transition from general to specific by layer-by-layer transformation \cite{yosinski2014transferable,zeiler2014visualizing}. But DSN breaks the procedure by direct supervision on earlier layers. Our method retains this procedure in deep nets, while yielding more variety in later layers.
 
 \subsection{Multi-branch Online Knowledge Distillation} Deep mutual learning \cite{zhang_deep_2018}, co-teaching \cite{han2018co} and co-distillation \cite{anil2018large} differ from previous methods. They enhance robustness or accuracy from training two or more networks simultaneously and let them teach each other during training. ONE \cite{lan2018knowledge} and CL-ILR \cite{song2018collaborative} construct auxiliary branches and let each branch learn from an ensemble teacher with a gate. OKDDip \cite{chen2020online} gains some performance from boosting peer diversity by introducing attention mechanism. The wisdom behind these methods resembles ensemble learning, but the difference is that every network will benefit from the knowledge taught by others during training, while ensemble learning works at inference time by gathering the knowledge together.
\section{Tree-structured auxiliary online knowledge distillation}
\label{method}

\subsection{A Unified Framework}
\label{unified}
Our TSA offers a unified perspective on designing architecture for online knowledge distillation. TSA dispenses with redesigning auxiliary modules, and leverages the original network block. We describe the framework as follows: for a target network, our TSA first splits the original network by depth into a few parts (i.e. network block), then adds more counterparts for ones close to the output layer, forming a tree structure. (Examples are shown in Fig. \ref{TSA example}.) Different initialization for each branch provide different view of the inputs and is the source of the knowledge transferred to each other. Sharing the same early layers not only cuts down the computational cost but also increases the performance of the overall architecture. We will discuss the underlying reasons in the following section. For a batch of data, inputs flow from the root to every leaf along all the branches simultaneously. Knowledge distillation is performed for outputs from every leaf. The whole tree is trained collaboratively to minimize a global loss during training. At inference time, auxiliary modules are removed to recover the architecture to the target network. However, if there is less constraint on computation budget and the performance is crucial, the ensemble model with all the auxiliary modules can be deployed.

Existing researches design architectures with auxiliary modules in a heuristic way. We propose a unified perspective here. DML\cite{zhang_deep_2018} can be interpreted as adding counterparts for every part of a network; ONE\cite{lan2018knowledge} and CL-ILR\cite{song2018collaborative} add counterparts only for later part in a network. TSA is a general framework, rather than any specific method. The overall architecture is determined with the configuration: which part of a network to duplicate, and the number of the duplication. It is flexible enough to be translated into different architecture suitable for specific scenario. Our TSA provides a general perspective to get better understanding of online knowledge distillation, and may inspire new effective training methods.

\begin{figure}
     \centering
     \begin{subfigure}[b]{0.48\linewidth}
         \centering
         \includegraphics[width=0.75\linewidth]{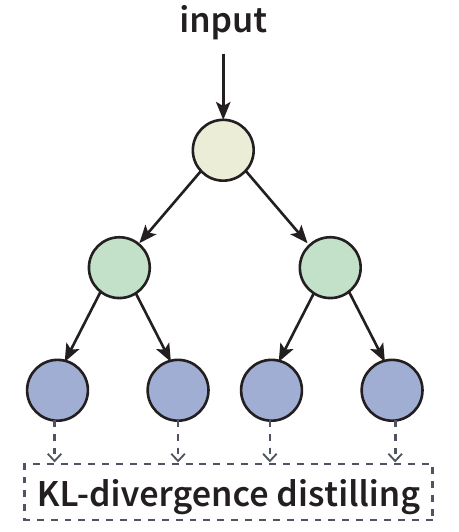}
         \caption{Balanced tree}
         \label{balanced tree}
     \end{subfigure}
     \begin{subfigure}[b]{0.49\linewidth}
         \centering
         \includegraphics[width=1\linewidth]{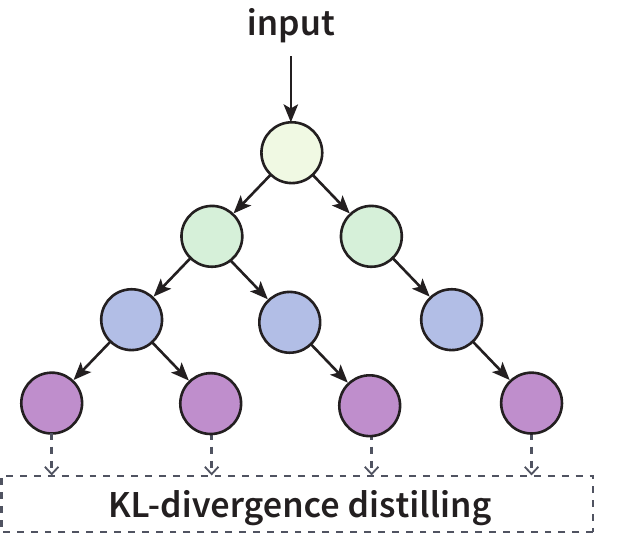}
         \caption{Unbalanced tree}
         \label{unbalanced tree}
     \end{subfigure}

        \caption{Two different settings of TSA. Left: balanced tree. Right: unbalanced tree. }
        \label{tree}
\end{figure}

Based on our framework, we propose two different settings of TSA. Fig. \ref{balanced tree} represents a balanced tree setting, Fig. \ref{unbalanced tree} represents an unbalanced one. Experiments show that the balanced tree setting has stronger regularization effect than the unbalanced one, thus yielding comparable or better performance. So we adopt the balanced tree as default setting of our TSA. 

TSA-M-H denotes a balanced tree of depth H, and internal node of the tree has M children. Thus this tree forms $M^{H-1}$ branches.
Fig. \ref{balanced tree} shows a case of TSA-2-3. We also explore trees with more depth and more children in Section \ref{component}, which requires more training cost but further boosts the performance. We adopt TSA-2-3 as our default setting for TSA.
\subsection{Formulation}

\paragraph{Softmax Function}
Given \textit{N} samples $X=\{x_i\}_{i=1}^{N}$ from $T$ classes, the corresponding label set is denoted as $Y=\{y_i\}_{i=1}^N, y_i\in\{1,2,\cdots,T\}$. We translate any original network into a tree structure by adding auxiliary duplicate modules. Classifier on every branch is denoted as $\Theta_k$, where $k\in\{1,2,\cdots,K\}$.
Classifier $\Theta_k$ produce class probabilities by applying the softmax:
\begin{equation}
    p_i=\frac{exp(z_i/T)}{\sum_{j}exp(z_j/T)},
\end{equation}
where the logit $z$ is the input to the final softmax, and $p_i$ is the $i^{th}$ class probability of classifier $\Theta_k$. T is temperature that is normally set to 1 to represent the cross-entropy. Higher value for T produces a softer probability distribution over class for distillation. \cite{hinton2015distilling}
\paragraph{Joint Loss Function}
Classifier $\Theta_k$ is trained with two loss terms. The first is a conventional cross-entropy loss to match the ground-truth labels:
\begin{equation}
    L_{C_k}=-\sum_{i,t}\delta_{i,t}\log{p_{i,t}},
\end{equation}
where $\delta_{i,t}=0$ if $y_i=t$, otherwise $\delta_{i,t}=1$.
This loss trains the network to predict correct labels for training datasets. The second loss is the Kullback Leibler (KL) Divergence to quantify the match of prediction $p_k$ to its peers. 
\begin{equation}
\label{kd_loss}
    L_{D_k}=\frac{1}{K-1}\sum_{j\neq{k}}KL(p_k || p_j)
\end{equation}
We perform distillation for every branch respectively, and this equation is in a symmetric form, although KL Divergence itself is not symmetric. The KL Divergence is calculated as below:
\begin{equation}
   KL(p_k || p_j)  = \sum_{j\neq{k}}\sum_{i,t}p_k^t(x_i) \log{\frac{p_k^t(x_i)}{p_j^t(x_i)}}
\end{equation}
 This loss enables knowledge distillation from peers to improve generalization performance on test sets. It is important to distil from each peer respectively, rather than distil from an average ensemble one. This provides enough variety and dispenses with extra effort on designing delicate mechanisms such as attention. We then obtain the overall loss for TSA as:
\begin{equation}
    L=\sum_k{[(1-\alpha)*L_{C_k}+\alpha*T^2*L_{D_k}]},
\end{equation}
where $\alpha$ is a hyperparameter to control the contribution of cross-entropy loss and distillation loss.

\subsection{Theoretical Insights}
Let $ h^l(x) \in R^d $ denote the output from the $l$-th layer of a neural network with input $x$. Then the output logit from the network is $q(y|x)=softmax(linear(h^L(...(h^1(x)))))$. We add more parallel later layers with different initialization with TSA. Let $h^l_i(x)$ denote the $i$-th peer, at layer $L$ , $i\in\{1,2,...,K\}$, which means the network has total K branches after adding auxiliary blocks. 

The TSA enhanced training can be formulated as the constrained optimization problem:
$$argmin_\xi(1/n*\sum_{i=1}^nE[-log(p(y_i|x_i))]),$$
$$s.t. \ \ (1/n)*\sum_{i=1}^nE_{\Theta_k,\Theta_t}[D_{KL}(p_k(y_i|x_i)||p_t(y_i|x_i))]=0,$$
where $\Theta_k,\Theta_t$ denote two different branch, and $\xi$ denotes the parameters of the network. The constraint from KL-divergence is equivalent to constraining output from different branches to be equal. This reduces the freedom of the parameters and the variance of individual solutions \cite{liang2021r}, forming a wider minimum from the view of optimization landscape \cite{zhang2019your,zhang_deep_2018}.

Since every branch share the same earlier layers in our method, for the shared parameters, they already satisfy the constraint in equation, which limit the KL-divergence loss to only take effect in optimization of the later layers.

Note that the development of deep learning theory is still at its infancy stage. And why knowledge distillation works is still in dispute \cite{phuong2019towards,allen2020towards}. We make some conjecture for the reason why the tree-structure boosts the performance of onine knowledge distillation based on our discussion above: earlier layers are extracting more general features, while layers close to the output are more task-specific. So only adding more branches for later layers improves the generalization to the target task, and leads to a broader optimization minimum. Different branches construct different views of the inputs, which can be the source of the knowledge. The hierarchical structure implies that the knowledge transfers from general to task-specific with the growth of the layers. We believe this is an important problem and leave more detailed analysis to our future work. 

\section{Experiments}
\label{experiments}

\begin{table*}
    \caption{Top-1 Accuracy (\%) on CIFAR-100 compared with other multi-branch methods. Our TSA adopts the TSA-2-3 version as Fig. \ref{balanced tree}. ``-E": ensemble result. The results reported for other methods are reproduced with the same setting. All the methods are with 4 branches for fair comparison. We report accuracy for single branch (without ``-E'') and 4-branch ensemble result. Our method outperforms other representative online knowledge distillation methods. }
      \label{cifar-100-table}
  \centering
    \begin{tabular}{lccccc}
    \toprule
    Method &  ResNet-32 & VGG-16 & ResNet-110 & MobileNetV3  & WRN-28-10\\
    \midrule
    Baseline    & 70.78 &  73.81    & 75.88      & 74.51        & 78.49 \\
    DML \cite{zhang_deep_2018}   & 74.01 & 75.44    & 76.33        & 72.56        & 80.28 \\

    ONE \cite{lan2018knowledge} & 73.84 & 74.50    & 77.72        & 76.11        & 81.67   \\
    OKDDip \cite{chen2020online}& 73.99 & 74.53    & 78.60        & 75.73        &80.88   \\
    TSA  &\textbf{74.57} &\textbf{76.58} & \textbf{79.75} & \textbf{78.93} &\textbf{82.13} \\
    \midrule
    ONE-E & 75.08 & 74.51 & 79.85 & 76.23 & 82.23 \\
    OKDDip-E & 75.74 & 74.44 & 79.62 & 77.47 & 81.82 \\
    TSA-E&\textbf{76.69}  &\textbf{78.28} & \textbf{81.96} &\textbf{79.62}& \textbf{83.03}\\
    \bottomrule
    \end{tabular}

\end{table*}

\begin{figure}
     \centering
     \includegraphics[width=2.3in]{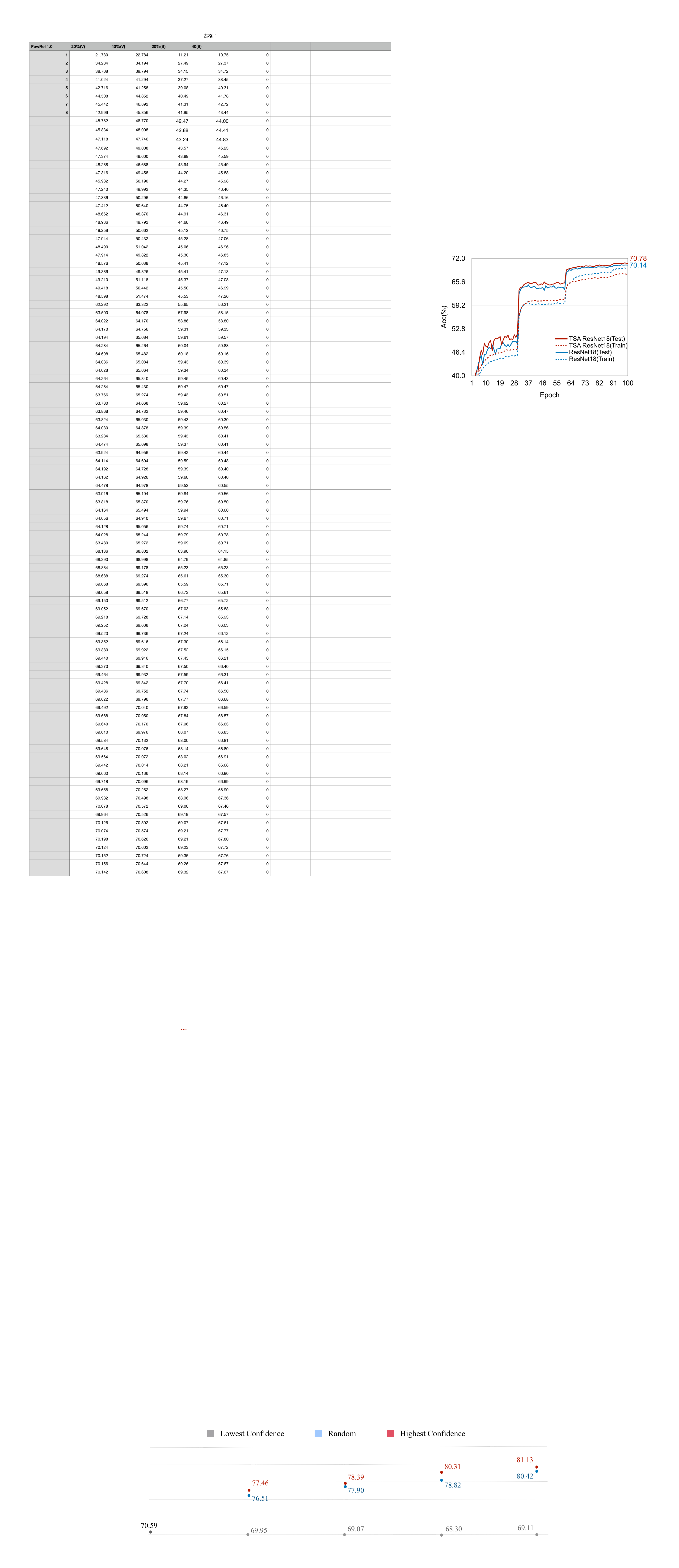}
     \caption{Training curve of top-1 Accuracy (\%) for ResNet-18 on ImageNet. We compare the performance of the original structure and the tree-structured model during training and testing.}
    \label{imagenet}

\end{figure}

\begin{table*}
     \caption{Top-1 Accuracy (\%) on CIFAR-100 compared with other deeply supervised methods. "Single": accuracy of a single classifier. "*": reported by \cite{zhang2019your}. Our method outperforms other methods by a large margin. U-TSA: unbalanced TSA showed in Fig. (\ref{unbalanced tree}).}
      \label{dsn-selfdistill-table}
  \centering
    \begin{tabular}{lcccccccc}
    \toprule
    \multirow{2}{*}[-2pt]{Model}& \multicolumn{4}{c}{Single} & \multicolumn{4}{c}{Ensemble}
    \\\cmidrule(lr){2-5}\cmidrule(lr){6-9}
               &DSN  & Self-distill & U-TSA & TSA   & DSN  & Self-distill & U-TSA & TSA \\\midrule
    VGG-19    & - & 68.03* & 73.62 & \textbf{74.03} & - & 68.54* & 74.83 & \textbf{75.94} \\
    ResNet-18   & 78.38* & 78.64* & 80.76 & \textbf{81.43}  & 79.27* & 79.67* & 82.13 & \textbf{82.43} \\\bottomrule
    \end{tabular}

\end{table*}

\subsection{Image Classification}
\label{cifar-100 section}
\paragraph{Datasets}
CIFAR-100 dataset \cite{krizhevsky2009learning} consists of $32 \times 32$ color images drawn from 100 classes split into 50000 train and 10000 test images. 
We apply standard data augmentation (translation and mirrorings) on this dataset.
ImageNet-1K 2012 \cite{deng2009imagenet} provides 1.2 million images for training, and 50k for validation. We report accuracy of ImageNet on the validation set, which is the standard process. We also include the experimental results on CIFAR-10 in a subsection.

\paragraph{Implementation Details} On CIFAR-100, we conduct all experiments on two NVIDIA GeForce GTX 1080 Ti GPUs. The initial learning rate is set to 0.1 and mini-batch size to 128. For ResNet and VGG, we adopt the same settings as \cite{lan2018knowledge} for fair comparisons. Learning rate drops by 0.1 at 50\% and 75\% training of total 300 training epochs. For MobileNetV3, we remove the first two pooling layers following \cite{haase2020rethinking} and train for 200 epochs with a weight decay of 0.0002. Learning rate decays by a factor of 0.1 at epochs 100, 150, and 180. For WideResNet, we follow \cite{zhang_deep_2018} to train for 200 epochs and learning rate drops by 0.1 every 60 epochs. On ImageNet, experiments are conducted on four NVIDIA GeForce GTX 3090 GPUs. We adopt the official code from PyTorch. \footnote{https://github.com/pytorch/examples/tree/master/imagenet} Initial learning rate is 0.1 and decay by 0.1 every 30 epochs for total of 100 epochs. We adopt the balanced tree setting as the default setting for TSA. We report the results with average 3 runs.

\paragraph{Comparison between the Tree Settings}
We compare the balanced tree setting and the unbalanced tree setting. We denote the unbalanced tree as "U-TSA". Table \ref{dsn-selfdistill-table} shows that TSA with balanced tree yields better performance than the unbalanced tree on CIFAR-100, however, on a larger dataset (ImageNet), these two settings have comparable results. (Shown in Table \ref{imagenet-table}.) This observation implies that the balanced tree has more regularizing effect due to the symmetric architecture. Large dataset alleviates the problem of overfitting, hence similar performance for these two settings. Without extra clarification, we adopt the balanced tree setting as the default setting below.

\begin{figure}
         \centering
         \includegraphics[width=2.3in]{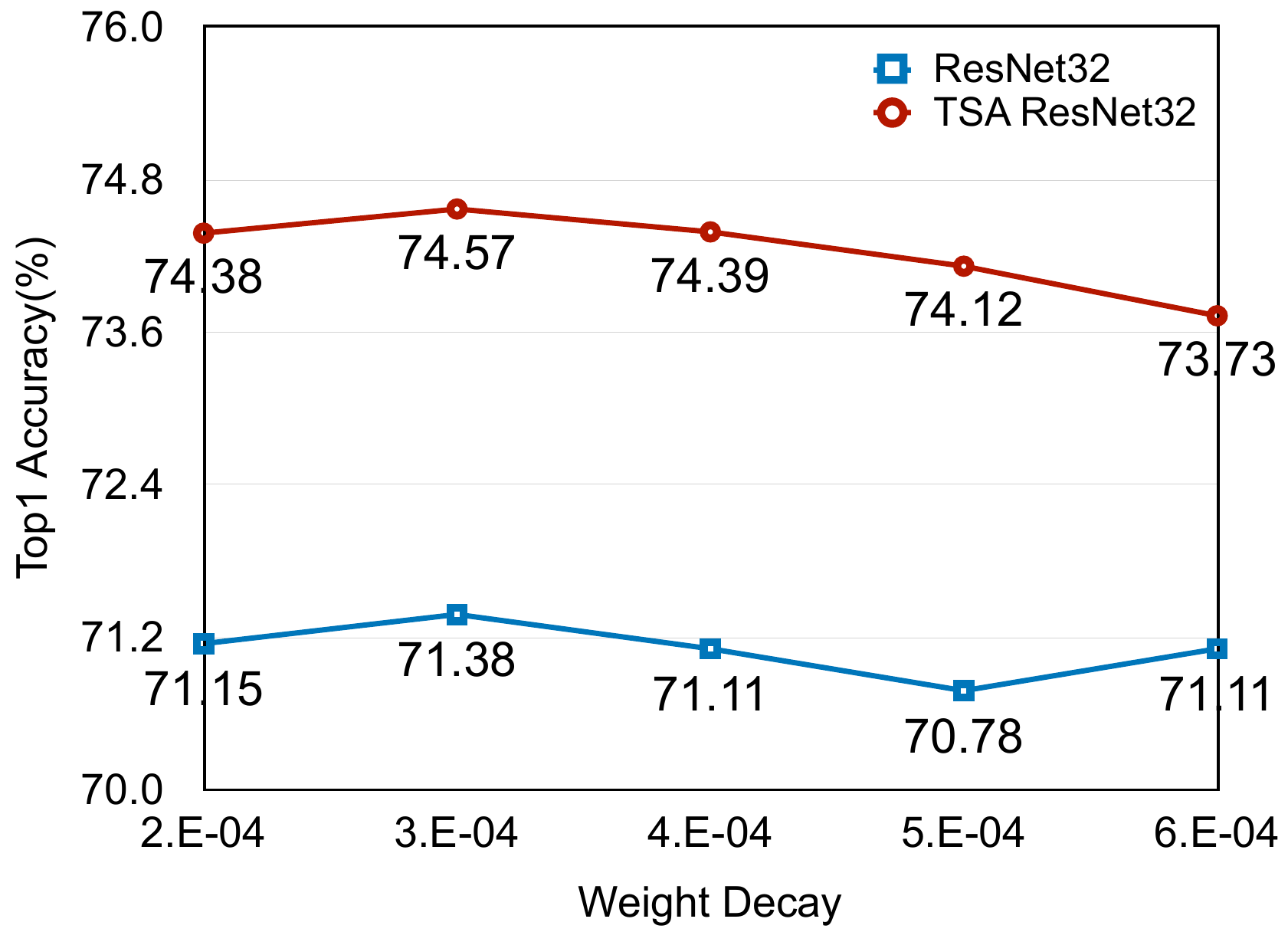}
         \caption{Top-1 Accuracy (\%) for ResNet-32 on CIFAR-100 with different weight decay. Our method is steadily above the baseline. }
         \label{resnet32-wd}
\end{figure}

\paragraph{Comparison with Vanilla Training}
Results are shown in Table \ref{cifar-100-table}. Note that in our method, we just keep the hyperparameters the same as the baseline and already get satisfying performance. Further tuning the hyperparameters may lead to further improvements. (See Fig. \ref{resnet32-wd}.) We evaluate our method with different networks, namely VGG-16 \cite{Simonyan15}, ResNet-32, ResNet-110 \cite{He_2016_CVPR}, MobileNetV3 \cite{howard2019searching}, and WRN-28-10 \cite{zagoruyko2016wide}. We use a large version for MobileNetV3. 

We focus on the accuracy boost of each method for target networks and ensemble networks. TSA gains an average of 3.52\% accuracy improvement over vanilla training. Improvements are consistent among non-residual network (VGG-16), very deep network (ResNet-110) and wide network (WRN-28-10). It is worth mentioning that on MobileNetV3, we achieve top-1 accuracy of 78.93\%, which is 1.23\% above the best number ever reported for this architecture. 

We observe that our method enhances more performance on models with complex architecture. ResNet-110 is deeper, WRN-28-10 is wider, and MobileNetV3 incorporates Squeeze-and-Excitation architecture which is more difficult to optimize. TSA enables these more complex architectures to converge to better minima. The same phenomenon is observed in Table \ref{imagenet-table}. ResNet-34 with TSA yields more significant improvement than ResNet-18 on ImageNet.

Fig. \ref{imagenet} shows the training and testing accuracy curve of ResNet18 with TSA and a vanilla one. We observe that the model yields higher test accuracy but lower training accuracy. This shows that our TSA acts as a regularizer, and knowledge distilling from peers still takes effect in the large dataset without extra knowledge.

\begin{table}
    \caption{Top-1 Accuracy (\%) on ImageNet. ``Single": result for a single branch target network. ``Ensemble": ensemble result for 4 branches. We simply average the outputs from these 4 branches as the ensemble.}
    \label{imagenet-table}
  \centering
    \begin{tabular}{lccccc}
    \toprule
    \multirow{2}{*}[-2pt]{Model}& \multicolumn{3}{c}{Single} & \multicolumn{2}{c}{Ensemble}
    \\\cmidrule(lr){2-4}\cmidrule(lr){5-6}
    &Vanilla & U-TSA & TSA & U-TSA & TSA\\
    \midrule
    ResNet18    & 70.14 & 70.80   & 70.74      & 71.91        & 72.13 \\
    ResNet34    & 73.24 & 74.93    & 75.03      & 76.22   & 76.47 \\
    \bottomrule
    \end{tabular}

\end{table}

\paragraph{Comparison with Multi-branch Methods}

We compare our TSA method with several representative multi-branch training methods, including DML \cite{zhang_deep_2018}, ONE \cite{lan2018knowledge} and OKDDip \cite{chen2020online}. 
Table \ref{cifar-100-table} reveals that our method outperforms other representative online knowledge distillation methods consistently. For fair comparison, we reproduce other methods with the same experimental setting and report the number for average 3 runs. Methods with "-E" means ensemble result. For ensemble results, all the methods are reported with 4-branch ensemble. Our TSA-E only simply averages the outputs from all the classifiers, but still outperforms other methods by a significant margin, which utilize complex components such as diversity and attention to get the ensemble model. This phenomenon implies that though we perform knowledge distillation between peers, some variety still exists in each branch.

\paragraph{Comparison with Deeply Supervised Methods}
We compare our TSA with DSN \cite{lee2015deeply} and self-distillation \cite{zhang2019your} for networks VGG-19 \cite{Simonyan15} and ResNet-18 \cite{He_2016_CVPR}. We adopt the same setting as \cite{zhang2019your} for a fair comparison. Results are shown in Table \ref{dsn-selfdistill-table}. Our TSA outperforms others by a large margin. It is worth mentioning that ResNet-18 achieves an 81.43\% accuracy with TSA, which even has higher accuracy than the vanilla residual network with over thousands of layers \cite{he2016identity}. Results of ResNet-18 (wider than ResNet-32) again show the favorable quality of our TSA, that TSA is more powerful when the model is larger or deeper.

\paragraph{Comparison to Ensemble of Baseline Network}
Table \ref{tab:ensemble} shows an example of comparison with ensemble results for vanilla network without auxiliary training. Note that TSA aims at training a single model to be more accurate and generalizable to the test data. That's why only single branch is left at test time. This resembles distilling knowledge from an ensemble model into a single one. So the idea is not to surpass an ensemble model, but to surpass the original model with the same parameter size. However, we still provide the comparison on CIFAR-100 in Table \ref{tab:ensemble} for better understanding of our method.

For ResNet-32, our method with single branch already outperforms 2-net ensemble of baseline model, and for MobileNetV3, it yileds similar performance with 4-net ensemble. Our TSA ensemble method yields best performance while having less parameters than vanilla 4-net ensemble due to the sharing of early layers.
\begin{table}[]
    \centering
        \caption{Top-1 accuracy of ResNet-32 on CIFAR-100}
    \label{tab:ensemble}
    \begin{tabular}{lcc}
    \toprule
     Method & ResNet-32 & MobileNetV3 \\
     \midrule
      Vanilla single net & 70.78 & 74.51 \\
      2-net ensemble & 73.42 & 77.34 \\
      3-net ensemble & 75.06 & 78.51 \\
      4-net ensemble & 75.69 & 78.97 \\
      TSA single branch & 74.12 & 78.93 \\
      TSA 4-branch ensemble & \textbf{76.69} & \textbf{79.62} \\
      \bottomrule
    \end{tabular}

\end{table}

\paragraph{Hyperparameter Sensitivity}
We train ResNet-32 with different weight decay values ranging from 0.0002 to 0.0006. Fig. \ref{resnet32-wd} shows that with different hyperparameters, the accuracy of ResNet-32 trained by TSA is steadily higher than the baseline in a relatively fixed range. That is to say, there is no need to pay great effort to search for the appropriate hyperparameters when applying our method. We also observe that the results reported in Table \ref{cifar-100-table} can be further improved, with more effort to find the best hyperparameter setting. In this case, with weight decay 0.0003, the accuracy 74.57\% is higher than that with the default setting.

\paragraph{Component Analysis}
\label{component}
Table \ref{component-table} shows component analysis results for different configuration for our TSA. TSA-M-H denotes training a model with an M-branch balanced tree of depth H. With distillation, the target model sees significant improvements. For a single target network, knowledge distillation facilitates the models to improve 1.98\% on average, compared with model without distillation. Specifically, TSA-2-4 performs best which suggests that TSA with more layers and peers distills fine-grained knowledge into each block. However, simple addition of branches gains little accuracy profit for "TSA-3-3".
Meanwhile, TSA without distillation performs better than baseline reported in Table \ref{cifar-100-table}, which implies that training all the sub-networks together yields more generalization.
For the ensemble networks, without constraints of distillation, more classifiers means higher accuracy.

\paragraph{CIFAR-10 Results for TSA}
Table \ref{tab:acc_cifar10} shows experiment results on CIFAR-10. For CIFAR-10 dataset, we follow the same experimental setting as CIFAR-100. We report comparison with former state-of-the-art method OKDDDip \cite{chen2020online}. Our method gains steady improvement upon baseline, while OKDDip suffers from unstability and sometimes sees performance decrease. The same problem is also observed for other auxiliary training methods in our experiments on CIFAR-10.
\begin{table}
    \centering
    \caption{Top-1 accuracy on CIFAR-10 dataset.}
    \label{tab:acc_cifar10}
    \begin{tabular}{lcccc}
    \toprule
     Model & Baseline & OKDDip & TSA & TSA-E \\
     \midrule
      ResNet-18 & 95.40 & - & 95.83 & \textbf{96.33} \\
      ResNet-32 & 93.66 & 93.48 & 94.47 & \textbf{95.08} \\
      MobileNetV3 & 92.40 & 91.79 & 94.39 & \textbf{94.63} \\
      ResNet-110 & 94.41 & 95.04 & 95.20 & \textbf{95.62} \\
      \bottomrule
    \end{tabular}

\end{table}

\begin{table}
  \centering
    \caption{Top-1 Accuracy (\%) for ResNet-32 on CIFAR-100. ``Single": accuracy of a single target network trained by TSA. ``w/ distil'' means that distillation is performed for outputs from different branch (See Equation \ref{kd_loss}), and ``w/o distil'' vice versa.}
      \label{component-table}
    \begin{tabular}{lccc}
    \toprule
    Configuration&TSA-2-3  & TSA-3-3 & TSA-2-4    \\
    \midrule
    Single w/o distil & 72.24 & 72.50 & 72.37 \\
    Ensemble w/o distil & 77.14 & \textbf{78.70} & 78.38 \\
    \midrule
    Single w/ distil & 74.12 & 74.17 & \textbf{74.65} \\
    Ensemble w/ distil  & 76.69 & 77.46 & 77.78 \\
    \bottomrule
    \end{tabular}
\end{table}

\subsection{Neural Machine Translation}
\label{sec::translation}
\paragraph{Datasets} We use 4 widely used dataset for translation task, including IWSLT'14 De-En~\cite{cettolo2014report}, IWSLT'17 \{De, Fr\}-En~\cite{cettolo2017overview} and WMT'17 En-De~\cite{tay2021omninet}. We use the first three datasets for small-scale translation tasks. And the larger WMT'17 dataset consists of about 4.5 million sentence pairs from WMT'14 \cite{vaswani2017attention}, with additional newstest14\footnote{http://nlp.stanford.edu/projects/nmt} data for test. We split the dataset into a shared source-target vocabulary of about 40000 tokens. 

\paragraph{Implementation Details} We train both original Transformer model and tree-structured Transformer model on one machine with 4 NVIDIA 3090 GPUs. The hyperparameter settings for both models are identical, using an initial learning rate of 0.0005 and 0.0007 for IWSLT series dataset and WMT'17 dataset, respectively. Each training batch contains a set of sentence pairs with approximately 25,000 source and 25,000 target tokens. The models for IWSLT series dataset are trained for 30 epochs, while each model for WMT'17 dataset is trained for a total of 80 epochs.
In terms of the model structure, each transformer encoder contains 6 identical encoder layers, so do the decoders. Our focus is on the validity of TSA in Transformer, but not on pushing the state-of-the-art results, so we simply adopt a 2-layer tree to speed up the training. We use half-precision floating-point numbers (FP16) during training.

\begin{figure}
     \centering
     \includegraphics[width=2.3in]{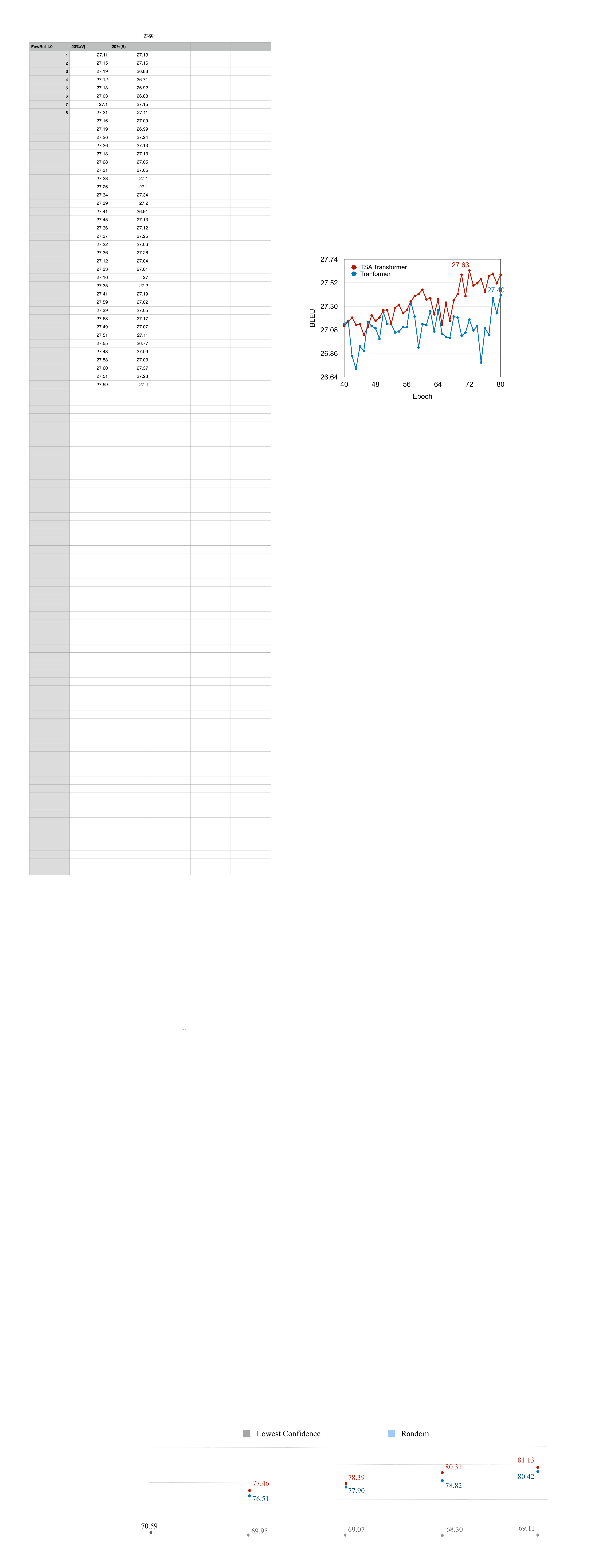}
     \caption{BLEU for Transformer on WMT'17. We report the performance of the original structure and tree-structured models in the last 40 training epochs. In training, TSA Transformer consistently outperforms original Transformer.}
         \label{translation}

\end{figure}

\paragraph{Comparison with Original Transformer} To show that TSA likewise boosts the performance of models for natural language processing tasks, we conduct comparative experiments on machine translation. Considering that Transformer is ubiquitous in natural language processing, and is extensively applied in computer vision recently, we compare the performance of the original Transformer with our tree-structured Transformer. We report the BLEU on the IWSLT series datasets and the results are shown in Table~\ref{tab:translation}. We observe that on these three small-scale datasets, TSA obtains a stable 
gain of about 0.9 BLEU on average, illustrating the effectiveness of TSA. Further, to demonstrate that TSA stabilizes the training process, we report the BLEU in the last 40 epochs on the WMT'17 dataset. As evident from Fig.~\ref{translation}, The BLEU curve of TSA shows a smooth and continuous increase, compared to the sharp fluctuations of the original transformer. During the training process, TSA Transformer performs steadily higher than the original Transformer, illustrating the robustness of our training method. It is promising that the performance will further improve if we train a tree with more layers. The experiments on Transformer confirm that our TSA is able to generalize from computer vision to natural language tasks.

\begin{table}[]
    \caption{BLEU for translation tasks. ``14": IWSLT'14 translation dataset; ``17'': IWSLT'17 translation dataset. We do not report results with other online knowledge distillation methods since none of previous methods have implemented application in Transformer. }
    \label{tab:translation}
    \centering
    \begin{tabular}{lccc}
    \toprule
     Method & 14 De-En & 17 De-En & 17 Fr-En\\
     \midrule
      Baseline & 34.4 & 27.6 & 36.6 \\
      TSA & 35.4 & 28.5 & 37.1 \\
      \bottomrule
    \end{tabular}
\end{table}

\subsection{Training Cost}
Table \ref{tab:para_size} shows the parameter size during training and Table \ref{tab:traing_time} shows training time for various networks. Our method slightly costs more than ONE and OKDDip for training. However, at deployment stage, since auxiliary branches are removed, the parameter size and inference time are all the same as baseline model. online knowledge distillation aims at training a better model with less deployment cost rather than training cost. For example, a vanilla ResNet-18 gets 78.18\% accuracy on CIFAR-100 while ResNet-101 79.29\%. With TSL, we train the ResNet-18 to achieve 81.43\% accuracy which is beyond vanilla ResNet-101, with far less parameters.

\begin{table}
    \centering
        \caption{Parameter size during training for different online knowledge distillation method.}
    \label{tab:para_size}
    \begin{tabular}{lcccc}
    \toprule
     Model & DML & ONE & OKDDip & TSA \\
     \midrule
      VGG-16 & 61.19M & 38.27M & 38.34M & 52.84M \\
      MobileNetV3 & 11.21M & 10.68M & 10.91M & 10.83M \\
      ResNet-32 & 1.89M & 1.55M & 1.55M & 1.64M \\
      ResNet-110 & 6.92M & 3.85M & 3.87M & 6.00M\\
      \bottomrule
    \end{tabular}

\end{table}

\begin{table}
    \centering
        \caption{Training time (seconds per epoch) on one Nvidia 2080 GPU on CIFAR-100 with batchsize is 128.}
    \label{tab:traing_time}
    \begin{tabular}{lcccc}
    \toprule
     Model & DML & ONE & OKDDip & TSA \\
     \midrule
      VGG-16 & 39 & 20 & 21 & 27 \\
      MobileNetV3 & 94 & 54 & 56 & 61 \\
      ResNet-32 & 33 & 23 & 23 & 28 \\
      ResNet-110 & 134 & 54 & 54 & 59\\
      \bottomrule
    \end{tabular}

\end{table}
\section{Discussion}
In this work, we show that through careful design of the overall architecture of the online knowledge distillation, our TSA achieves state-of-the-art performance with the vanilla online distillation loss function.  We introduce a unified framework for architecture design of online knowledge distillation and propose two different settings of TSA and demonstrates the superiority of our method over existing methods. Experiments on both image classification tasks and translation task certify the effectiveness of TSA. We are the first to demonstrate the effectiveness of online knowledge distillation for neural machine translation task. Our method is simple and effective, and it is promising that introducing advanced training objective \cite{lan2018knowledge,chen2020online} will further improve the performance of TSA. We leave that to our future work.
\bibliographystyle{plain}
\bibliography{ref}

\end{document}